\documentclass[12pt]{article}
\usepackage[margin=1in]{geometry}
\usepackage{amsthm} 
\usepackage{setspace} 
\usepackage{amsmath, amssymb, amsfonts, graphicx, color, bm, float, bbm, natbib}
\usepackage{cases} 
\usepackage{hyperref}
\newtheorem{theorem}{Theorem}

\newcommand{\myurl}[2]{\url{#1}} 
\newcommand{\x}{\bm{x}}
\newcommand{\y}{\bm{y}}
\newcommand{\s}{\bm{s}}
\newcommand{\yobs}{\bm{y}_{\text{obs}}}
\newcommand{\sobs}{\bm{s}_{\text{obs}}}
\newcommand{\ABC}{\text{ABC}}

\DeclareMathOperator{\Unif}{Unif}
\floatstyle{ruled}
\newfloat{Algorithm}{htbp}{alg}

\title{Adapting the ABC distance function}

\author{Dennis Prangle\footnote{Newcastle University.  Email \href{mailto:dennis.prangle@gmail.com}{\nolinkurl{dennis.prangle@gmail.com}}}}
\date{}

\begin{document}

\maketitle

\begin{abstract}
Approximate Bayesian computation performs approximate inference for models where likelihood computations are expensive or impossible.
Instead simulations from the model are performed for various parameter values and accepted if they are close enough to the observations.
There has been much progress on deciding which summary statistics of the data should be used to judge closeness, but less work on how to weight them.
Typically weights are chosen at the start of the algorithm which normalise the summary statistics to vary on similar scales.
However these may not be appropriate in iterative ABC algorithms, where the distribution from which the parameters are proposed is updated.
This can substantially alter the resulting distribution of summary statistics, so that different weights are needed for normalisation.
This paper presents two iterative ABC algorithms which adaptively update their weights and demonstrates improved results on test applications.
\end{abstract}

\section{Introduction}


Approximate Bayesian computation (ABC) is a family of approximate inference methods which can be used when the likelihood function is expensive or impossible to compute but simulation from the model is straightforward. 
The simplest algorithm is a form of rejection sampling.
Here parameter values are drawn from the prior distribution and corresponding datasets simulated.
Each simulation is converted to a vector of summary statistics $\s=(s_1, s_2, \ldots, s_m)$ and a distance between this and the summary statistics of the observed data, $\sobs$, is calculated.
Parameters producing distances below some threshold are accepted and form a sample from an approximation to the posterior distribution.

The choice of summary statistics has long been recognised as being crucial to the quality of the approximation \citep{Beaumont:2002}, but there has been less work on the role of the distance function.
A popular distance function is weighted Euclidean distance:
\begin{equation} \label{eq:WeightedEuclideanDist}
d(\s,\sobs) = \left[ \sum_{i=1}^m \left( \frac{s_i-s_{\text{obs}, i}}{\sigma_i} \right)^2 \right]^{1/2}
\end{equation}
where $\sigma_i$ is an estimate of the prior predictive standard deviation of the $i$th summary statistic.
In ABC rejection sampling a convenient estimate is the empirical standard deviation of the simulated $s_i$ values.
Scaling by $\sigma_i$ in \eqref{eq:WeightedEuclideanDist} normalises the summaries so that they vary over roughly the same scale,
preventing the distance being dominated by the most variable summary.

This paper concerns the choice of distance in more efficient iterative ABC algorithms, in particular those of \cite{Toni:2009}, \cite{Sisson:2009} and \cite{Beaumont:2009}.
The first iteration of these algorithms is the ABC rejection sampling algorithm outlined above.
The sample of accepted parameters is used to construct an importance density.
An ABC version of importance sampling is then performed.
This is similar to ABC rejection sampling, except parameters are sampled from the importance density rather than the prior, and the output sample is weighted appropriately to take this change into account.
The idea is to concentrate computational resources on performing simulations for parameter values likely to produce good matches.
The output of this step is used to produce a new importance density and perform another iteration, and so on.
In each iteration the acceptance threshold is reduced, resulting in increasingly accurate approximations.
Full details of the \cite{Toni:2009} implementation are reviewed later.

Weighted Euclidean distance is commonly used in these algorithms with $\sigma_i$ values determined in the first iteration.
However there is no guarantee that these will normalise the summary statistics produced in later iterations, as these are no longer drawn from the prior predictive.
This paper proposes two variant iterative ABC algorithms which update their $\sigma_i$ values to appropriate values at each iteration.
It is demonstrated that these algorithms provide substantial advantages in applications.
Also, they do not require any extra simulations to be performed solely for tuning.
Therefore even when a non-adaptive distance performs adequately, there is no major penalty in using the new approach.
(Some additional calculations are required -- calculating more $\sigma_i$ values and more expensive distance calculations -- but these form a negligible part of the overall computational cost.)

One of the proposed algorithms has similarities to the iterative ABC methods of \cite{Sedki:2012} and \cite{Bonassi:2015}.
These postpone deciding some elements of the tuning of iteration $t$ until during that iteration.
Algorithm \ref{alg:adaptive2} also uses this strategy but for different tuning decisions: the distance function and the acceptance threshold.
Another related paper is \cite{Fasiolo:2015} which contains an illustration of the difficulty of choosing ABC distance weights non-adaptively.

The remainder of the paper is structured as follows.
Section \ref{sec:ABC} reviews ABC algorithms.
This includes some novel material on the convergence of iterative ABC methods.
Section \ref{sec:distances} discusses weighting summary statistics in a particular ABC distance function.
Section \ref{sec:methods} details the proposed algorithms.
Several examples are given in Section \ref{sec:examples}.
Section \ref{sec:conclusion} summarises the work and discusses potential extensions.
Finally Appendix \ref{sec:convergence} contains technical material on convergence of ABC algorithms.
Computer code to implement the methods of this paper in the Julia programming language \citep{Bezanson:2012} is available at \myurl{https://github.com/dennisprangle/ABCDistances.jl}{https://github.com/\allowbreak dennisprangle/ABCDistances.jl}.

\section{Approximate Bayesian Computation} \label{sec:ABC}


This section sets out the necessary background on ABC algorithms.
Several review papers \citep[e.g.][]{Beaumont:2010, Csillery:2010, Marin:2012} give detailed descriptions of other aspects of ABC, including tuning choices and further algorithms.
Sections \ref{sec:ABCrej} and \ref{sec:ABCPMC} review ABC versions of rejection sampling and PMC.
Section \ref{sec:ABCPMCconvI} contains novel material on the convergence of ABC algorithms.

\subsection{ABC rejection sampling} \label{sec:ABCrej}

Consider Bayesian inference for parameter vector $\theta$ under a model with density $\pi(\y|\theta)$.
Let $\pi(\theta)$ be the prior density and $\yobs$ represent the observed data.
It is assumed that $\pi(\y|\theta)$ cannot easily be evaluated but that it is straightforward to sample from the model.
ABC rejection sampling (Algorithm \ref{alg:ABCrej}) exploits this to sample from an approximation to the posterior density $\pi(\theta|\y)$.
It requires several tuning choices:
number of simulations $N$,
a threshold $h \geq 0$,
a function $S(\y)$ mapping data to a vector of summary statistics,
and a distance function $d(\cdot,\cdot)$.

\begin{Algorithm}
\caption{ABC-rejection}
\label{alg:ABCrej}
\begin{enumerate}
\item Sample $\theta^*_i$ from $\pi(\theta)$ independently for $1 \leq i \leq N$.
\item Sample $\y^*_i$ from $\pi(\y|\theta^*_i)$ independently for $1 \leq i \leq N$.
\item Calculate $\s^*_i=S(\y^*_i)$ for $1 \leq i \leq N$.
\item Calculate $d^*_i = d(\s^*_i, \sobs)$ (where $\sobs=S(\yobs)$.)
\item Return $\{ \theta^*_i | d^*_i \leq h \}$.
\end{enumerate}
\end{Algorithm}

The threshold $h$ may be specified in advance.
Alternatively it can be calculated following step 4.
For example a common choice is to specify an integer $k$ and take $h$ to be the $k$th smallest of the $d_i^*$ values \citep{Biau:2015}.

\subsection{ABC-PMC} \label{sec:ABCPMC}

Algorithm \ref{alg:ABCPMC} is an iterative ABC algorithm taken from \cite{Toni:2009}.
Very similar algorithms were also proposed by \cite{Sisson:2009} and \cite{Beaumont:2009}.
The latter note that this approach is an ABC version of population Monte Carlo \citep{Cappe:2004}, so it is referred to here as ABC-PMC.
The algorithm involves a sequence of thresholds, $(h_t)_{t \geq 1}$.
Similarly to $h$ in ABC-rejection, this can be specified in advance or during the algorithm, as discussed below.

\begin{Algorithm}
\caption{ABC-PMC (with the option of adaptive $h_t$)}
\label{alg:ABCPMC}
\begin{enumerate}
\item[] {\bf Initialisation}
\item Let $t=1$.
\item[] {\bf Main loop}
\item Repeat following steps until there are $N$ acceptances.
\begin{enumerate}
\item Sample $\theta^*$ from importance density $q_t(\theta)$ given in equation \eqref{eq:qt}.
\item If $\pi(\theta^*)=0$ reject and return to (a).
\item Sample $\y^*$ from $\pi(\y|\theta^*_i)$ and calculate $\s^*=S(y^*)$.
\item Accept if $d(\s^*, \sobs) \leq  h_t$.
\end{enumerate}
\item[] Denote the accepted parameters as $\theta^t_1, \ldots, \theta^t_N$ and the corresponding distances as $d^t_1, \ldots, d^t_N$.
\item Let $w_i^t = \pi(\theta_i^t)/q_t(\theta_i^t)$ for $1 \leq i \leq N$.
\item (Optional) Let $h_{t+1}$ be the $\alpha$ quantile of the $d^t_i$ values.
\item Increment $t$ to $t+1$.
\item[] {\bf End of loop}
\end{enumerate}
\end{Algorithm}

The algorithm samples parameters from the importance density
\begin{subnumcases}{q_t(\theta) = \label{eq:qt}}
\pi(\theta) & if $t=1$, or $t=2$ and $h_1=\infty$ \\
\sum_{i=1}^N w_i^{t-1} K_t(\theta | \theta_i^{t-1}) / \sum_{i=1}^N w_i^{t-1} & otherwise. \label{eq:qtb}
\end{subnumcases}
In the first iteration (and sometimes the second, as discussed shortly) $q_t(\theta)$ is the prior.
Otherwise \eqref{eq:qtb} is used, which effectively samples from the previous weighted population and perturbs the result using kernel $K_t$.
\cite{Beaumont:2009} show that a good choice of the latter is
\[
K_t(\theta | \theta') = \phi(\theta', 2 \Sigma_{t-1}),
\]
where $\phi$ is the density of a normal distribution and $\Sigma_{t-1}$ is the empirical variance matrix of $(\theta_i^{t-1})_{1 \leq i \leq N}$ calculated using weights $(w_i^{t-1})_{1 \leq i \leq N}$

As mentioned above, the sequence of thresholds can be specified in advance.
However it is hard to do this well.
A popular alternative \citep{Drovandi:2011macroparasite} is to choose the thresholds adaptively by setting $h_t$ at the end of iteration $t-1$ to be the $\alpha$ quantile of the accepted distances (n.b.~$\alpha<1$ is assumed throughout the paper).
An optional step, step 4, is included in Algorithm \ref{alg:ABCPMC} to implement this method.
Alternative updating rules for $h_t$ have been proposed such as choosing it to reduce an estimate of effective sample size by a prespecified proportion \citep{delMoral:2012} or using properties of the predicted ABC acceptance rate \citep{Silk:2013}.

If step 4 is used this leaves $h_1$ and $\alpha$ as tuning choices.
A simple default for $h_1$ is $\infty$, in which case all simulations are accepted when $t=1$.
In this case \eqref{eq:qtb} would give $q_2(\theta)$ as simply a modified prior with inflated variance, which is not a sensible importance density.
Therefore \eqref{eq:qt} takes $q_2(\theta) = \pi(\theta)$ in this case.
This is a minor novelty of this presentation of the algorithm.

A practical implementation of Algorithm \ref{alg:ABCPMC} requires a condition for when to terminate.
In this paper the total number of datasets to simulate is specified as a tuning parameter and the algorithm stops once a further simulation is required.
Some alternative are possible, such as stopping once the algorithm falls below a target value for $h_t$ or the acceptance rate.

Several variations on Algorithm \ref{alg:ABCPMC} have been proposed which are briefly discussed in Section \ref{sec:conclusion}.
Some of these are ABC versions of sequential Monte Carlo (SMC).
The phrase ``iterative ABC'' will be used to cover ABC-PMC and ABC-SMC.

\subsection{Convergence of ABC-PMC} \label{sec:ABCPMCconvI}

Conditions C1-C5 ensure that Algorithm \ref{alg:ABCPMC} converges on the posterior density in an appropriate sense as the number of iterations tends to infinity.
This follows from Theorem \ref{thm:convergence} which is described in Appendix \ref{sec:convergence}.
Although only finite computational budgets are available in practice, such convergence at least guarantees that the target distribution become arbitrarily accurate as computational resources are increased.

\begin{itemize}
\item[C1.]
$\theta \in \mathbb{R}^n$, $\s \in \mathbb{R}^m$ for some $m,n$ and these random variables have density $\pi(\theta, \s)$ with respect to Lebesgue measure.
\item[C2.]
The sets $A_t = \{ \s | d(\s,\sobs) \leq h_t \}$ are Lebesgue measurable.
\item[C3.]
$\pi(\sobs) > 0$.
\item[C4.]
$\lim_{t \to \infty} |A_t| = 0$ (where $|\cdot|$ represents Lebesgue measure.)
\item[C5.]
The sets $A_t$ have \emph{bounded eccentricity}.
\end{itemize}
Bounded eccentricity is defined in Appendix \ref{sec:convergence}.
Roughly speaking, it requires that under any projection of $A_t$ to a lower dimensional space the measure still converges to zero.

Condition C1 is quite strong, ruling out discrete parameters and summary statistics, but makes proof of Theorem \ref{thm:convergence} straightforward.
Condition C2 is a mild technical requirement.
The other conditions provide insight into conditions required for convergence.
Condition C3 requires that it must be possible to simulate $\sobs$ under the model.
Condition C4 requires that the acceptance regions $A_t$ shrink to zero measure.
For most distance functions this corresponds to $\lim_{t \to \infty} h_t = 0$.
It is possible for this to fail.
Some examples encountered by the author in practice follow.
One is when datasets close to $\sobs$ cannot be produced under the model of interest.
Alternatively, even if $\sobs$ can occur under the model, the algorithm may converge on importance densities on $\theta$ under which it is impossible.
This corresponds to concentrating on the wrong mode of the ABC target distribution in an early iteration.
Finally, condition C5 prevents $A_t$ converging to a set where some but not all summary statistics are perfectly matched.

Conditions C4 and C5 can be used to check which distance functions are sensible to use in ABC-PMC, usually by investigating whether they hold when $h_t \to 0$.
For example it is straightforward to show this is the case when $d(\cdot,\cdot)$ is a metric induced by a norm.

\section{Weighted Euclidean distance in ABC} \label{sec:distances}


This paper concentrates on using weighted Euclidean distance in ABC.
Section \ref{sec:Euclidean} discusses this distance and how to choose its weights.
Section \ref{sec:scale_illustration} illustrates its usefulness in a simple example.

\subsection{Definition and usage} \label{sec:Euclidean}

Consider the following distance:
\begin{equation} \label{eq:weighted_d}
d(\x,\y) = \left[ \sum_{i=1}^m \left\{ \omega_i (x_i - y_i) \right\}^2 \right]^{1/2}.
\end{equation}
If $\omega_i=1$ for all $i$, this is is \emph{Euclidean distance}.
Otherwise it is a form of \emph{weighted Euclidean distance}.

Many other distance functions can be used in ABC,
as discussed in Section \ref{sec:ABCPMCconvI},
for example weighted $L_1$ distance
$d(\x,\y) = \sum_{i=1}^m \omega_i |x_i - y_i|$.
To the author's knowledge the only published comparison of distance functions is by \cite{McKinley:2009}, which found little difference between the alternatives.
\cite{Owen:2015} report the same conclusion but not the details.
This finding is also supported in unpublished work by the author of this paper.
Given these empirical results this paper focuses on \eqref{eq:weighted_d} as it is a simple choice, but no claims are made for its optimality.
Some further discussion on this is given in Section \ref{sec:conclusion}.

Summary statistics used in ABC may vary on substantially different scales.
In the extreme case Euclidean distance will be dominated by the most variable.
To avoid this, weighted Euclidean distance is generally used.
This usually takes $\omega_i = 1/\sigma_i$ where $\sigma_i$ is an estimate of the scale of the $i$th summary statistic.
(Using this choice in weighted Euclidean distance gives the distance function \eqref{eq:WeightedEuclideanDist} discussed in the introduction.)

A popular choice  \citep[e.g.][]{Beaumont:2002} of $\sigma_i$ is the empirical standard deviation of the $i$th summary statistic under the prior predictive distribution.
\cite{Csillery:2012} suggest using median absolute deviation (MAD) instead since it is more robust to large outliers.
MAD is used throughout this paper.
For many ABC algorithms these $\sigma_i$ values can be calculated without requiring any extra simulations.
For example this can be done between steps 3 and 4 of ABC-rejection.
ABC-PMC can be modified similarly, resulting in Algorithm \ref{alg:ABCPMC2}, which also updates $h_t$ adaptively.
(n.b.~All of the ABC-PMC convergence discussion in Section \ref{sec:ABCPMCconvI} also applies to this modification.)


\begin{Algorithm}
\caption{ABC-PMC with adaptive $h_t$ and $d(\cdot,\cdot)$}
\label{alg:ABCPMC2}
\begin{enumerate}
\item[] {\bf Initialisation}
\item Let $t=1$ and $h_1=\infty$.
\item[] {\bf Main loop}
\item Repeat following steps until there are $N$ acceptances.
\begin{enumerate}
\item Sample $\theta^*$ from importance density $q_t(\theta)$ given in equation \eqref{eq:qt}.
\item If $\pi(\theta^*)=0$ reject and return to (a).
\item Sample $\y^*$ from $\pi(\y|\theta^*_i)$ and calculate $\s^*=S(y^*)$.
\item Accept if $d(\s^*, \sobs) \leq  h_t$ (if $t=1$ always accept).
\end{enumerate}
\item If $t=1$:
\begin{enumerate}
\item Calculate $(\sigma_1, \sigma_2, \ldots)$, a vector of MADs for each summary statistic, calculated from all the simulations in step 2 (including those rejected).
\item Define $d(\cdot,\cdot)$ as the distance \eqref{eq:weighted_d} using weights $(\omega_i)_{1 \leq i \leq m}$ where $\omega_i = 1/\sigma_i$.
\end{enumerate}
\item[] Denote the accepted parameters as $\theta^t_1, \ldots, \theta^t_N$ and the corresponding distances as $d^t_1, \ldots, d^t_N$.
\item Let $w_i^t = \pi(\theta_i^t)/q_t(\theta_i^t)$ for $1 \leq i \leq N$.
\item Let $h_t$ be the $\alpha$ quantile of the $d^t_i$ values.
\item Increment $t$ to $t+1$.
\item[] {\bf End of loop}
\end{enumerate}
\end{Algorithm}

\subsection{Illustration} \label{sec:scale_illustration}

\begin{figure*}[htb]
\begin{center}
\includegraphics{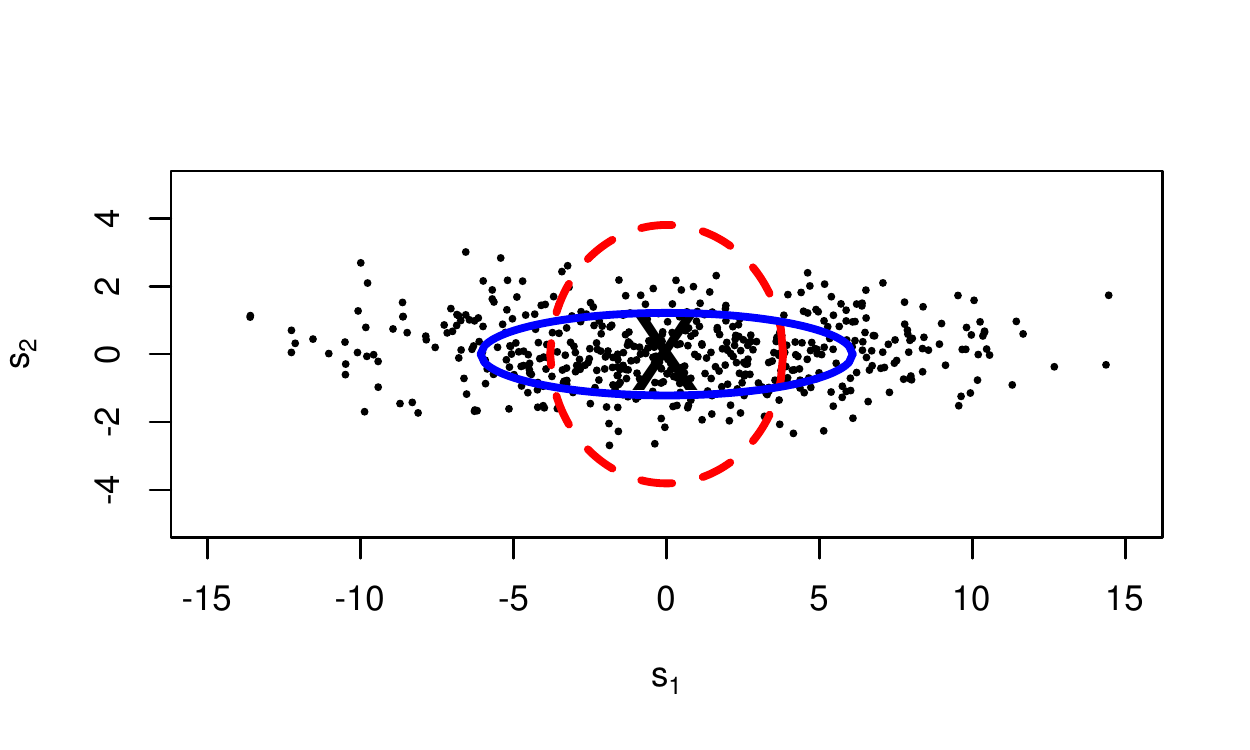}
\end{center}
\caption{An illustration of distance functions in ABC rejection sampling.
The points show simulated summary statistics $s_1$ and $s_2$.
The observed summary statistics are taken to be $(0,0)$ (black cross).
Acceptance regions are shown for two distance functions, Euclidean (red dashed circle) and weighted Euclidean with MAD reciprocals as weights (blue solid ellipse).
These show the sets within which summaries are accepted.
The acceptance thresholds have been tuned so that each region contains half the points.
}
\label{fig:dist_illustration}
\end{figure*}

As an illustration, Figure \ref{fig:dist_illustration} shows the difference between using Euclidean and weighted Euclidean distance with $\omega_i=1/\sigma_i$ within ABC-rejection.
Here $\sigma_i$ is calculated using MAD.
For both distances the acceptance threshold is tuned to accept half the simulations.
In this example Euclidean distance mainly rejects simulations where $s_1$ is far from its observed value: it is dominated by this summary.
Weighted Euclidean distance also rejects simulations where $s_2$ is far from its observed value and is less stringent about $s_1$.

Which of these distances is preferable depends on the relationship between the summaries and the parameters.
For example if $s_1$ were the only informative summary, then Euclidean distance would preferable.
In practice, this relationship may not be known.
Weighted Euclidean distance is then a sensible choice as both summary statistics contribute to the acceptance decision.

This heuristic argument supports the use of weighted Euclidean distance in ABC more generally.
One particular case is when low dimensional informative summary statistics have been selected, for example by the methods reviewed in \cite{Blum:2013}.
In this situation all summaries are known to be informative and should contribute to the acceptance decision.

Note that in Figure \ref{fig:dist_illustration} the observed summaries $\sobs$ lie close to the centre of the set of simulations.
When some observed summaries are hard to match by model simulations this is not the case.
ABC distances could now be dominated by the summaries which are hardest to match.
How to weight summaries in this situation is discussed in Section \ref{sec:conclusion}.

\section{Methods: Iterative ABC with an adaptive distance} \label{sec:methods}


The previous section discussed normalising ABC summary statistics using estimates of their scale under the prior predictive distribution.
This prevents any summary statistic dominating the acceptance decision in ABC-rejection or the first iteration of Algorithm \ref{alg:ABCPMC2}, where the simulations are generated from the prior predictive.
However in later iterations of Algorithm \ref{alg:ABCPMC2} the simulations may be generated from a very different distribution so that this scaling is no longer appropriate.
This section presents two versions of ABC-PMC which avoid this problem by updating the distance function at each iteration.
Normalisation is now based on the distribution of summary statistics generated in the previous (Algorithm \ref{alg:adaptive1}) or current (Algorithm \ref{alg:adaptive2}) iteration.
The proposed algorithms are presented in Sections \ref{sec:ABCadaptive1} and \ref{sec:ABCadaptive2}.

An approach along these lines has the danger that the summary statistic acceptance regions at each iteration no longer form a nested sequence of subsets converging on the point $\s=\sobs$.
To avoid this, the proposed algorithms only accept a simulated dataset at iteration $t$ if it also meets the acceptance criteria of \emph{every previous iteration}.
This can be viewed as sometimes modifying the $t$th distance function to take into account information from previous iterations.
Section \ref{sec:ABCPMCconvII} discusses convergence in more depth.

\subsection{First proposed algorithm} \label{sec:ABCadaptive1}

Algorithm \ref{alg:adaptive1} is a straightforward modification of Algorithm \ref{alg:ABCPMC2} which updates its distance function at each iteration using scales derived from the previous iteration's simulations.
The first iteration accepts everything so no distance function is required.
This acts as an initial tuning step.
Note that scales are based on both accepted and rejected simulations from the previous iteration.
This is because using just the accepted simulations would mean the scales are sometimes mainly determined by the previous acceptance rule, restricting the scope for adaptation.

\begin{Algorithm}
\caption{ABC-PMC with adaptive $h_t$ and $d^t(\cdot,\cdot)$}
\label{alg:adaptive1}
\begin{enumerate}
\item[] {\bf Initialisation}
\item Let $t=1$ and $h_1=\infty$.
\item[] {\bf Main loop}
\item Repeat following steps until there are $N$ acceptances.
\begin{enumerate}
\item Sample $\theta^*$ from importance density $q_t(\theta)$ given in equation \eqref{eq:qt}.
\item If $\pi(\theta^*)=0$ reject and return to (a).
\item Sample $\y^*$ from $\pi(\y|\theta^*_i)$ and calculate $\s^*=S(y^*)$.
\item If $t=1$ accept. Otherwise accept if $d^i(\s^*,\sobs) \leq h_i$ for all $2 \leq i \leq t$.
\end{enumerate}
\item Calculate $(\sigma_1^t, \sigma_2^t, \ldots)$, a vector of MADs for each summary statistic, calculated from all the simulations in step 2 (including those rejected).
\item Define $d^{t+1}(\cdot,\cdot)$ as the distance \eqref{eq:weighted_d} using weights $(\omega_i)_{1 \leq i \leq m}$ where $\omega_i = 1/\sigma_i$.
\item[] Denote the accepted parameters as $\theta^t_1, \ldots, \theta^t_N$ and the corresponding distances under $d^{t+1}(\cdot, \cdot)$ as $d^{t+1}_1, \ldots, d^{t+1}_N$.
\item Let $w_i^t = \pi(\theta_i^t)/q_t(\theta_i^t)$ for $1 \leq i \leq N$.
\item Let $h_{t+1}$ be the $\alpha$ quantile of the $d^{t+1}_i$ values.
\item Increment $t$ to $t+1$.
\item[] {\bf End of loop}
\end{enumerate}
\end{Algorithm}

Storing all simulated $\s^*$ vectors to calculate scale estimates in step 3 of Algorithm \ref{alg:adaptive1} can be impractical.
In practice storage is stopped after the first few thousand simulations, and scale estimation is done using this subset.
Other tuning details of Algorithm \ref{alg:adaptive1}
-- the choice of perturbation kernel $K_t$ and the rule to terminate the algorithm --
are implemented as described earlier for ABC-PMC.

\subsection{Second proposed algorithm} \label{sec:ABCadaptive2}

Algorithm \ref{alg:adaptive1} normalises simulations in iteration $t$ based on scales derived in the preceding iteration.
This could be inappropriate if two consecutive iterations sometimes generate simulations from markedly different distributions. 
Algorithm \ref{alg:adaptive2} addresses this problem.

A naive approach would be to start iteration $t$ by tuning $d^t(\cdot, \cdot)$ using an additional set of simulations based on parameters drawn from the current importance distribution.
However this imposes an additional cost.
Instead the algorithm makes a single large set of simulations.
These are first used to construct the $t$th distance function.
Then the best $N$ simulations are accepted and used to construct the next importance distribution.

A complication is deciding how many simulations to make for this large set.
There must be enough that $N$ of them are accepted.
However the distance function defining the acceptance rule is not known until after the simulations are performed.
The solution implemented is to continue simulating until $M=\lceil N/\alpha \rceil$ simulations pass the acceptance rule of the \emph{previous} iteration.
Let $\mathcal{A}$ be the set of these simulations and $\mathcal{B}$ be the others.
Next the new distance function is constructed (based on $\mathcal{A} \cup \mathcal{B}$) and the $N$ with lowest distances (from $\mathcal{A}$) are accepted.
The tuning parameter $\alpha$ has a similar interpretation to the corresponding parameter in Algorithms \ref{alg:ABCPMC2} and \ref{alg:adaptive1}:
the acceptance threshold in iteration $t$ is the $\alpha$ quantile of the realised distances from simulations in $\mathcal{A}$.

Using this approach means that, as well as adapting the distance function, another difference with Algorithms \ref{alg:ABCPMC2} and \ref{alg:adaptive1} is that selection of $h_t$ is delayed from the end of iteration $t-1$ to part-way through iteration $t$ (and therefore $h_1$ does not need to be specified as a tuning choice.)
If desired, this novelty can be used without adapting the distance function.
Such a variant of Algorithm \ref{alg:ABCPMC2} was tried on the examples of this paper, but the results are omitted as performance is very similar to Algorithm \ref{alg:ABCPMC2}.

Given the same importance density and acceptance rule, an iteration of Algorithm \ref{alg:adaptive2} requires the same expected number of simulations as Algorithms \ref{alg:ABCPMC2} and \ref{alg:adaptive1}.
In this sense their costs are the same.
In practice, the algorithms select their importance density and acceptance rules differently so this comparison of their computational costs is limited.
Section \ref{sec:examples} contains empirical comparisons in terms of the mean squared error for a given number of simulations.

\begin{Algorithm}
\caption{ABC-PMC with adaptive $h_t$ and $d^t(\cdot,\cdot)$}
\label{alg:adaptive2}
\begin{enumerate}
\item[] {\bf Initialisation}
\item Let $t=1$.
\item[] {\bf Main loop}
\item Repeat following steps until there are $M=\lceil N/\alpha \rceil$ acceptances.
\begin{enumerate}
\item Sample $\theta^*$ from importance density $q_t(\theta)$ given in equation \eqref{eq:qt}.
\item If $\pi(\theta^*)=0$ reject and return to (a).
\item Sample $\y^*$ from $\pi(\y|\theta^*_i)$ and calculate $\s^*=S(y^*)$.
\item If $t=1$ accept. Otherwise accept if $d^i(\s^*,\sobs) \leq h_i$ for all $i<t$.
\end{enumerate}
\item[] Denote the accepted parameters as $\theta^*_1, \ldots, \theta^*_M$ and the corresponding summary vectors as $\s^*_1, \ldots, \s^*_M$.
\item Calculate $(\sigma_1^t, \sigma_2^t, \ldots)$, a vector of MADs for each summary statistic, calculated from all the simulations in step 2 (including those rejected).
\item Define $d^t(\cdot,\cdot)$ as the distance \eqref{eq:weighted_d} using weights $(\omega^t_i)_{1 \leq i \leq m}$ where $\omega^t_i = 1/\sigma^t_i$.
\item Calculate $d^*_i = d^t(\s^*_i, \sobs)$ for $1 \leq i \leq M$.
\item Let $h_t$ be the $N$th smallest $d^*_i$ value.
\item Let $(\theta_i^t)_{1 \leq i \leq N}$ be the $\theta_i^*$ vectors with the smallest $d^*_i$ values (breaking ties randomly).
\item Let $w_i^t = \pi(\theta^t_i)/q_t(\theta^t_i)$ for $1 \leq i \leq N$.
\item Increment $t$ to $t+1$.
\item[] {\bf End of loop}
\end{enumerate}
\end{Algorithm}

The comments at the end of Section \ref{sec:ABCadaptive1} on tuning details and storing $\s^*$ vectors also apply to Algorithm \ref{alg:adaptive2}.


\subsection{Convergence} \label{sec:ABCPMCconvII}

This section shows that conditions for the convergence of Algorithms \ref{alg:adaptive1} and \ref{alg:adaptive2} in practice are essentially those described in Section \ref{sec:ABCPMCconvI} for standard ABC-PMC plus one extra requirement:
$e_t = \frac{\max_i w^t_i}{\min_i w^t_i}$ is bounded above.

In more detail, conditions ensuring convergence of Algorithms \ref{alg:adaptive1} and \ref{alg:adaptive2} can be taken from Theorem \ref{thm:convergence} in Appendix \ref{sec:convergence}.
These are the same as those given for other ABC-PMC algorithms in Section \ref{sec:ABCPMCconvI} with the exception that the acceptance region $A_t$ is now defined as
$\{ \s | d^i(\s,\sobs) \leq h_i \text{ for all } i \leq t \}$.
Two conditions behave differently under this change: C4 and C5.

Condition C4 states that $\lim_{t \to \infty} |A_t| = 0$ i.e.~Lebesgue measure tends to zero.
The definition of $A_t$ for Algorithms \ref{alg:adaptive1} and \ref{alg:adaptive2} ensures $|A_t|$ is decreasing in $t$.
However it may not converge to zero.
Reasons for this are the same as why condition C4 can fail for standard ABC-PMC, as described in Section \ref{sec:ABCPMCconvI}.

Condition C5 is bounded eccentricity (defined in Appendix \ref{sec:convergence}) of the $A_t$ sets.
Under distance \eqref{eq:weighted_d} this can easily be seen to correspond to $e_t$ having an upper bound.
This is not guaranteed by Algorithms \ref{alg:adaptive1} and \ref{alg:adaptive2}, but it can be imposed, for example by updating $\omega^t_i$ to $\omega^t_i + \delta \max_i{\omega^t_i}$ after step 4 for some small $\delta>0$.
However this was not found to be necessary in any of the examples of this paper.

\section{Examples} \label{sec:examples}


This section presents three examples comparing the proposed and existing ABC-PMC algorithms:
a simple illustrative normal model, the $g$-and-$k$ distribution and the Lotka-Volterra model.

\subsection{Normal distribution} \label{sec:illustration_normal}

Suppose there is a single parameter $\theta$ with prior distribution $N(0,100^2)$.
Let $s_1 \sim N(\theta, 0.1^2)$ and $s_2 \sim N(0, 1^2)$ independently.
These are respectively informative and uninformative summary statistics.
Let $s_{\text{obs}, 1} = s_{\text{obs}, 2} = 0$.

\begin{figure*}[pht]
\includegraphics[width=\textwidth]{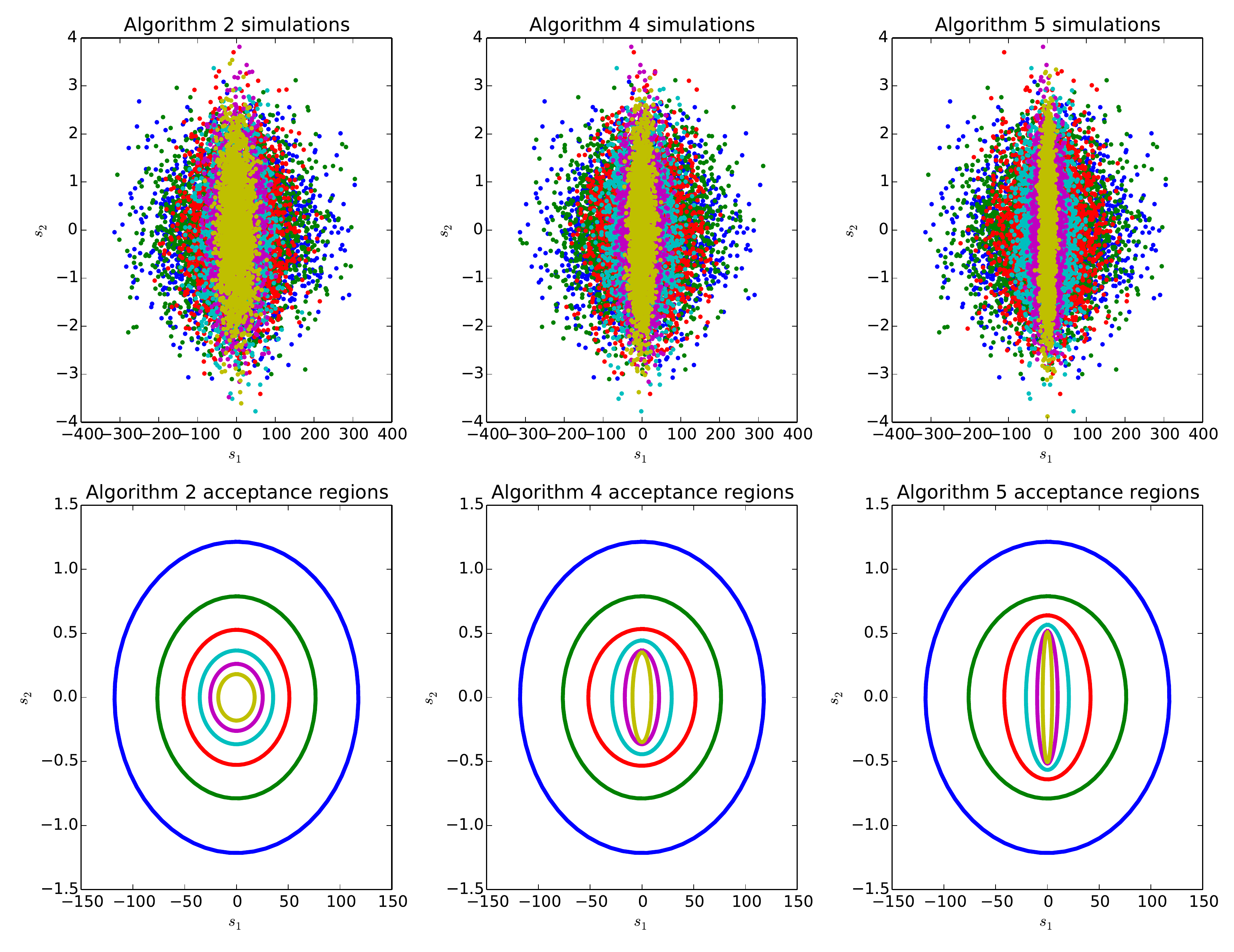}
\caption{An illustration of ABC-PMC for a simple normal model using Algorithms \ref{alg:ABCPMC} (non-adaptive distance function), \ref{alg:adaptive1} and \ref{alg:adaptive2} (adaptive distance functions).
\emph{Top row:} simulated summary statistics (including rejections)
\emph{Bottom row:} acceptance regions (note different scale to top row).
In both rows colour indicates the iteration of the algorithm.
}
\label{fig:normal_illustration}
\end{figure*}

\begin{figure*}[pht]
\includegraphics[width=\textwidth]{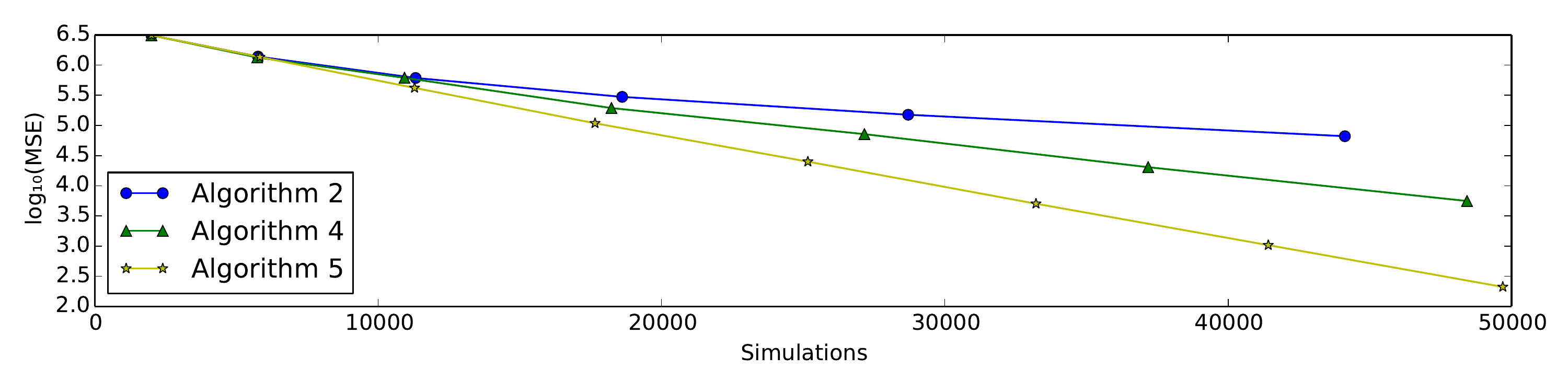}
\caption{Mean squared error of the parameter for a simple normal example using Algorithms \ref{alg:ABCPMC}, \ref{alg:adaptive1} and \ref{alg:adaptive2}.}
\label{fig:normal_RMSE}
\end{figure*}

Figures \ref{fig:normal_illustration} and \ref{fig:normal_RMSE} illustrate the behaviour of ABC-PMC for this example using Algorithms \ref{alg:ABCPMC} (with adaptive choice of $h_t$), \ref{alg:adaptive1} and \ref{alg:adaptive2}.
For ease of comparison the algorithms use the same random seed,
and the distance function and first threshold value $h_1$ for Algorithms \ref{alg:ABCPMC} and \ref{alg:adaptive1} are specified to be those produced in the first iteration of Algorithm \ref{alg:adaptive2}.
The effect is similar to making a short preliminary run of ABC-rejection to make these tuning choices.
All algorithms use $N=2000$ and $\alpha=1/2$.
(Empirical tests show that $\alpha \approx 1/2$ minimises mean squared error for all algorithms in this and the following examples.)

Under the prior predictive distribution the MAD for $s_1$ is in the order of 100 while that for $s_2$ is in the order of 1.
Therefore the first acceptance region in Figure \ref{fig:normal_illustration} is a wide ellipse.
Under Algorithm \ref{alg:ABCPMC} the subsequent acceptance regions are smaller ellipses with the same shape and centre.
The acceptance regions for Algorithms \ref{alg:adaptive1} and \ref{alg:adaptive2} are similar for the first few iterations.
After this, enough has been learnt about $\theta$ that the simulated summary statistics have a different distribution, with a reduced MAD for $s_1$.
Hence $s_1$ is given a larger weight, while the MAD and weight of $s_2$ remain roughly unchanged.
Thus the acceptance regions change shape to become narrower ellipses,
which results in a more accurate estimation of $\theta$, as shown by the comparison of mean squared errors (MSEs) in Figure \ref{fig:normal_RMSE}.
Note that Algorithm \ref{alg:adaptive2} adapts its weights more quickly than Algorithm \ref{alg:adaptive1} and hence achieves a smaller MSE.

\subsection{$g$-and-$k$ distribution}

The $g$-and-$k$ distribution is a popular test of ABC methods.
It is defined by its quantile function:
\begin{equation} \label{eq:gk}
A + B \left[ 1 + c\frac{1-\exp(-gz(x))}{1+\exp(-gz(x))} \right] [1+z(x)^2]^k z(x),
\end{equation}
where $z(x)$ is the quantile function of the standard normal distribution.
Following the literature \citep{Rayner:2002}, $c=0.8$ is used throughout.
This leaves $(A,B,g,k)$ as unknown parameters.

The $g$-and-$k$ distribution does not have a closed form density function making likelihood-based inference difficult.
However simulation is straightforward:
sample $x \sim \Unif(0,1)$ and substitute into \eqref{eq:gk}.
The following example is taken from \cite{Drovandi:2011replenishment}.
Suppose a dataset is 10,000 independent identically distributed draws from the $g$-and-$k$ distribution and the summary statistics are a subset of the order statistics: those with indices $(1250,2500,\ldots,8750)$.
(As in \citealp{Fearnhead:2012}, a fast method is used to simulate these order statistics without sampling an entire dataset.)
The parameters are taken to have independent $\Unif(0,10)$ priors.

To use as observations, 100 datasets are simulated from the prior predictive distribution.
Each is analysed using Algorithms \ref{alg:ABCPMC2}, \ref{alg:adaptive1} and \ref{alg:adaptive2}.
All analyses uses a total of $10^6$ simulations and tuning parameters $N=1000$ and $\alpha=1/2$.
Table \ref{tab:gkRMSE} shows root mean squared errors for the output of the algorithms, averaged over all the observed datasets.
These show that the adaptive algorithms, \ref{alg:adaptive1} and \ref{alg:adaptive2}, are more accurate overall for every parameter, and perform very similarly to each other.

\begin{table}[hptb]
\begin{center}
\begin{tabular}{c|cccc}
                              & A     & B     & g     & k     \\
\hline
Algorithm \ref{alg:ABCPMC2}   & 0.335 & 0.501 & 0.880 & 0.163 \\
Algorithm \ref{alg:adaptive1} & 0.083 & 0.371 & 0.532 & 0.126 \\
Algorithm \ref{alg:adaptive2} & 0.081 & 0.373 & 0.523 & 0.126
\end{tabular}
\end{center}
\caption{Root mean squared errors of each parameter in the $g$-and-$k$ example, averaged over analyses of 100 simulated datasets.} \label{tab:gkRMSE}
\end{table}

\begin{figure*}[p]
\includegraphics[width=\textwidth]{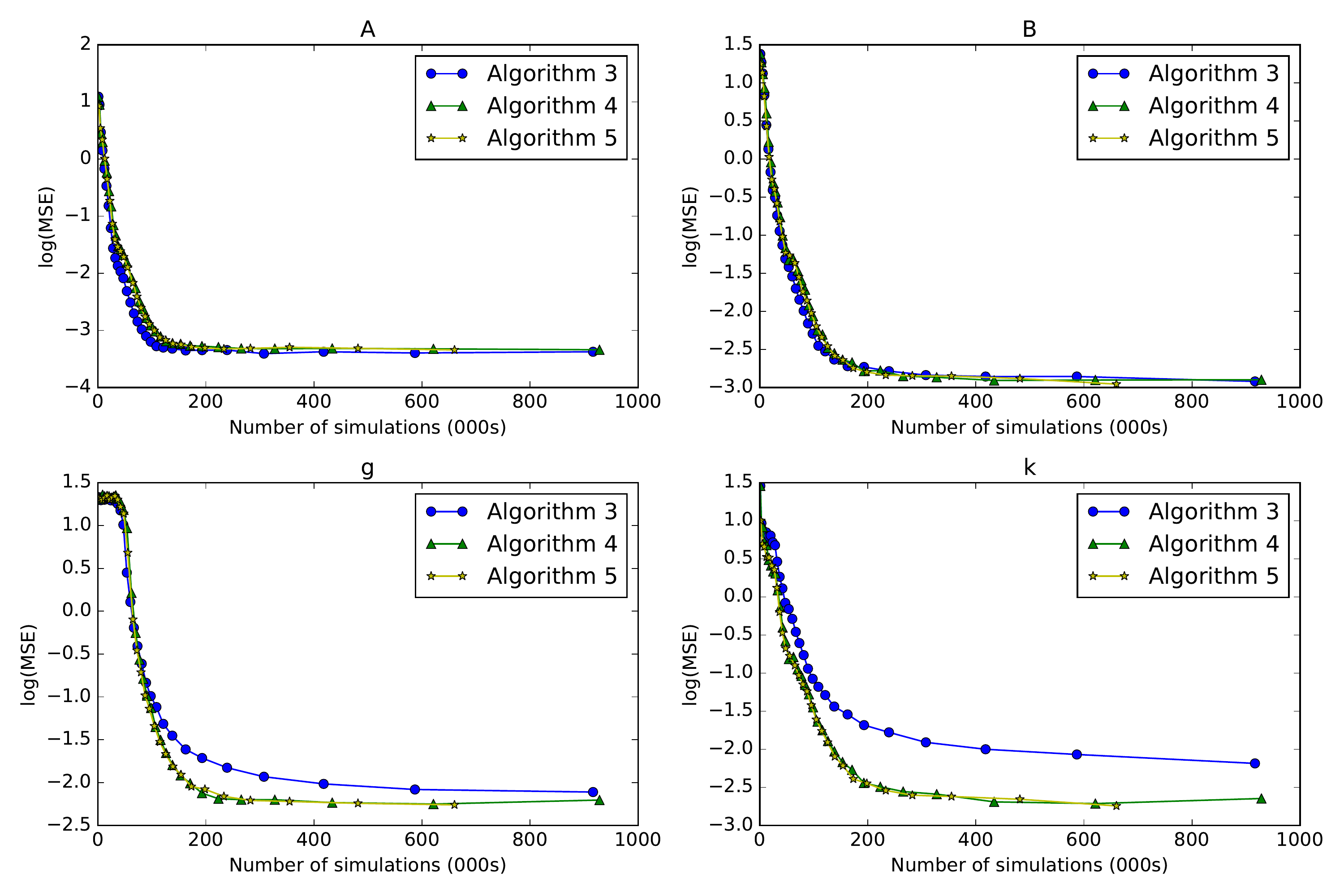}
\caption{Mean squared error of each parameter from Algorithms \ref{alg:ABCPMC2}, \ref{alg:adaptive1} and \ref{alg:adaptive2} for the $g$-and-$k$ example.}
\label{fig:gkbiasvar}
\end{figure*}

More detail is now given for a particular observed dataset, simulated under parameter values $(3, 1, 1.5, 0.5)$.
Figure \ref{fig:gkbiasvar} shows the estimated MSE of each parameter for each iteration of the three algorithms.
The adaptive algorithms, \ref{alg:adaptive1} and \ref{alg:adaptive2}, performs better throughout for the $g$ and $k$ parameters.
For this dataset all the algorithms perform similarly for the location and scale parameters $A$ and $B$, which have smaller MSE values.
Table \ref{tab:gkpost} demonstrates that the main difference in the final estimated posteriors is that Algorithm \ref{alg:ABCPMC2} has higher variances for the $g$ and $k$ parameters.

\begin{table}[hptb]
\begin{center}
\begin{tabular}{c|cccc}
                              & A            & B            & g            & k            \\
\hline
Algorithm \ref{alg:ABCPMC2}   & 2.98 (0.012) & 0.98 (0.028) & 1.52 (0.086) & 0.50 (0.081) \\
Algorithm \ref{alg:adaptive1} & 2.98 (0.012) & 0.97 (0.025) & 1.56 (0.048) & 0.53 (0.035) \\
Algorithm \ref{alg:adaptive2} & 2.98 (0.012) & 0.98 (0.024) & 1.56 (0.046) & 0.53 (0.033)
\end{tabular}
\end{center}
\caption{Estimated marginal posterior means and standard deviations (in brackets) of each parameter in the $g$-and-$k$ example, for analysis of a particular simulated dataset.
The values are taken from the final iteration of each algorithm.
(n.b.~All the estimated posteriors are roughly normal.)
} \label{tab:gkpost}
\end{table}

\begin{figure*}[p]
\includegraphics[width=\textwidth]{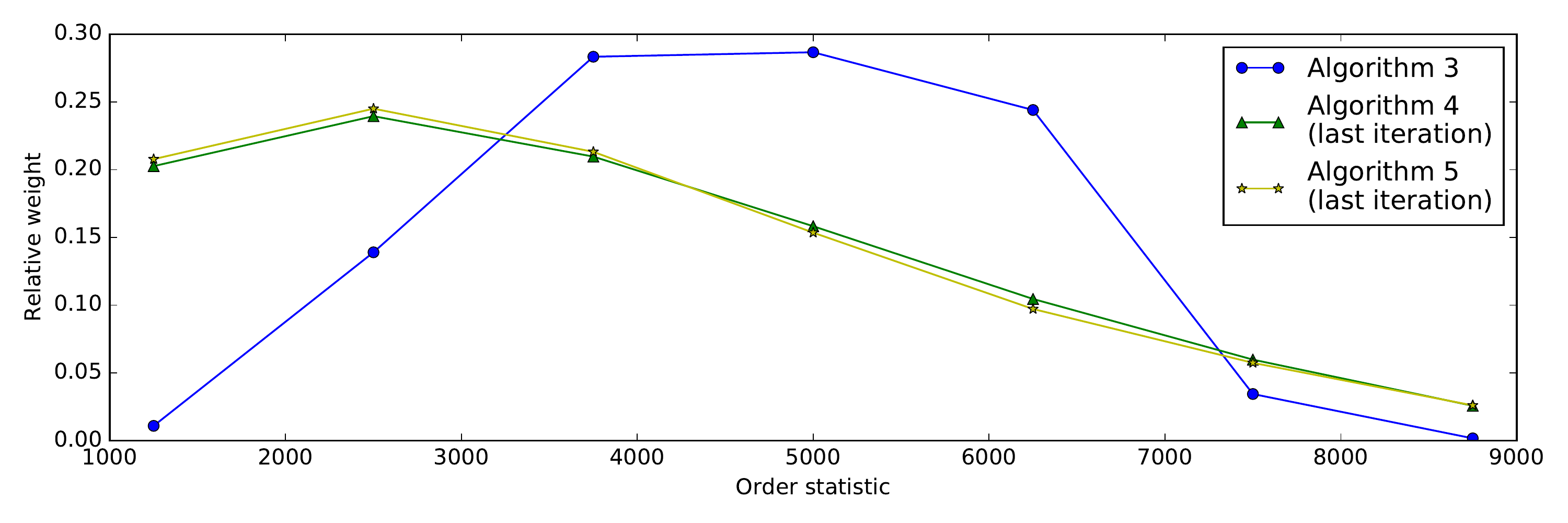}
\caption{Summary statistic weights used in Algorithms \ref{alg:ABCPMC2}, \ref{alg:adaptive1} and \ref{alg:adaptive2} for the $g$-and-$k$ example, rescaled to sum to 1.
}
\label{fig:gkweights}
\end{figure*}

Figure \ref{fig:gkweights} shows some of the distance function weights produced by the algorithms.
Algorithm \ref{alg:ABCPMC2} places low weights on the most extreme order statistics, as they are highly variable in the prior predictive distribution.
This is because the prior places significant weight upon parameter values producing very heavy tails.
However by the last iteration of Algorithms \ref{alg:adaptive1} and \ref{alg:adaptive2} such parameter values have been ruled out.
The algorithm therefore assigns larger weights which provide access to the informational content of these statistics.

\subsection{Lotka-Volterra model}

The Lotka-Volterra model describes two interacting populations.
In its original ecological setting the populations represent predators and prey.
However it is also a simple example of biochemical reaction dynamics of the kind studied in systems biology.
This section concentrates on a stochastic Markov jump process version of this model with state $(X_1,X_2) \in \mathbb{Z}^2$ representing prey and predator population sizes.
Three transitions are possible:
\begin{center}
\begin{tabular}{ll}
$(X_1,X_2) \to (X_1+1, X_2)$   & (prey growth)    \\
$(X_1,X_2) \to (X_1-1, X_2+1)$ & (predation)      \\
$(X_1,X_2) \to (X_1, X_2-1)$   & (predator death) \\
\end{tabular}
\end{center}
These have hazard rates $\theta_1 X_1$, $\theta_2 X_1 X_2$ and $\theta_3 X_2$ respectively.
Simulation is straightforward by the Gillespie method.
Following either a transition at time $t$, or initiation at $t=0$, the time to the next transition is exponentially distributed with rate equal to the sum of the hazard rates at time $t$.
The type of the next transition has a multinomial distribution with probabilities proportional to the hazard rates.
For more background see for example \cite{Owen:2015}, from which the following specific inference problem is taken.

The initial conditions are taken to be $X_1=50, X_2=100$.
A dataset is formed of observations at times $2, 4, 6, \ldots, 32$.
Both $X_1$ and $X_2$ are observed plus independent $N(0,\sigma^2)$ errors, where $\sigma$ is fixed at $\exp(2.3)$.
The unknown parameters are taken to be $\log \theta_1, \log \theta_2$ and $\log \theta_3$.
These are given independent $\Unif(-6,2)$ priors.
The vector of all 32 noisy observations is used as the ABC summary statistics.

A single simulated dataset is analysed (shown in Figure \ref{fig:LVpaths}.)
This is generated from the model with $\theta_1=1, \theta_2=0.005, \theta_3=0.6$.
ABC analysis is performed using Algorithms \ref{alg:ABCPMC2}, \ref{alg:adaptive1} and \ref{alg:adaptive2}.
A total of $50,000$ simulations are used in each algorithm.
The tuning parameters are $N=200$ and $\alpha=1/2$.
Any Lotka-Volterra simulation reaching $100,000$ transitions is terminated and automatically rejected.
This avoids extremely long simulations, such as exponential prey growth if predators die out.
These incomplete simulations are excluded from the MAD calculations, but this should have little effect as they are rare.

\begin{figure*}[p]
\includegraphics[width=\textwidth]{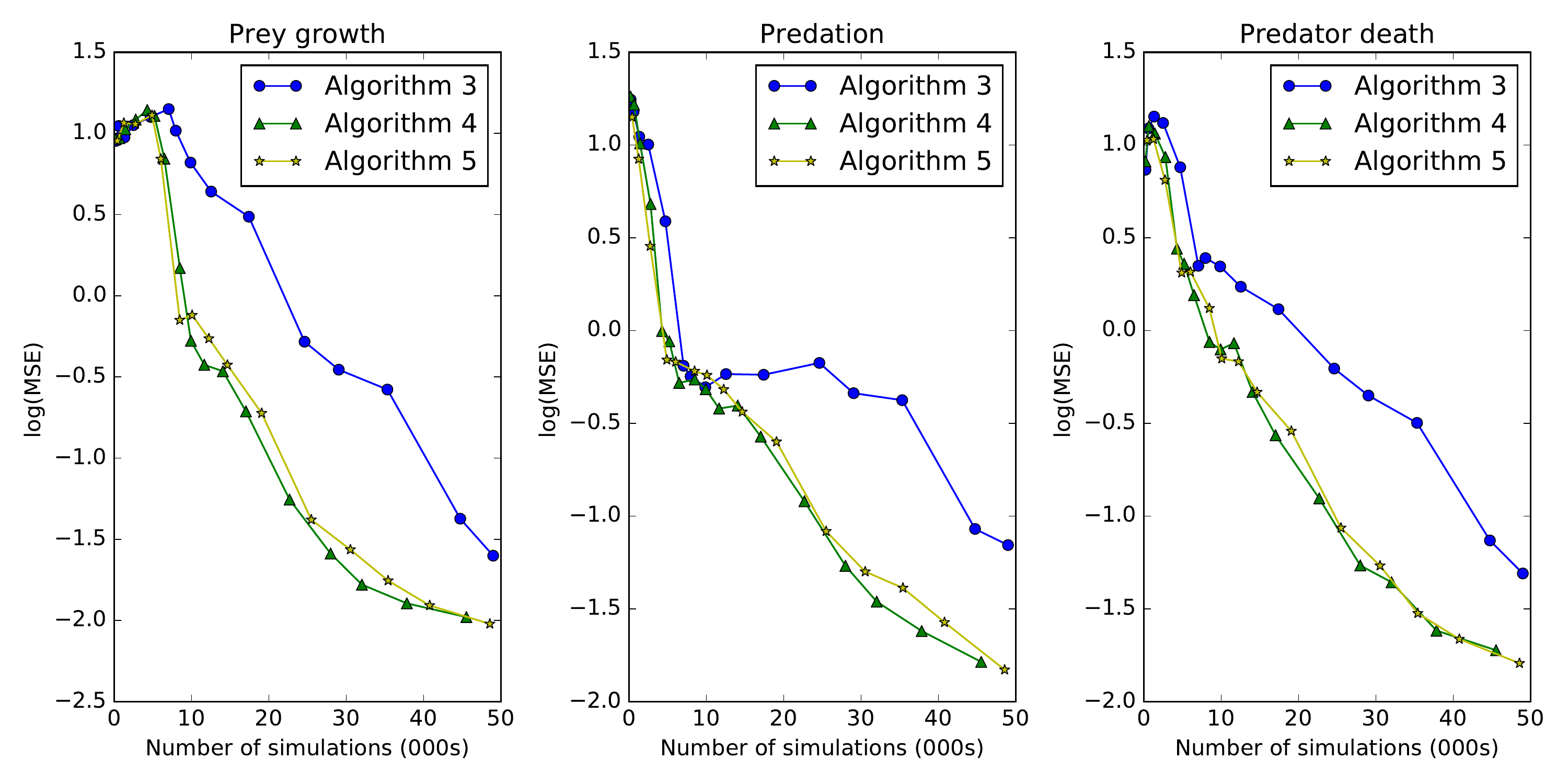}
\caption{Mean squared error of each parameter (i.e.~$\log \theta_1, \log \theta_2, \log \theta_3$) from ABC-PMC output for the Lotka-Volterra example.}
\label{fig:LVmse}
\end{figure*}

\begin{figure*}[p]
\includegraphics[width=\textwidth]{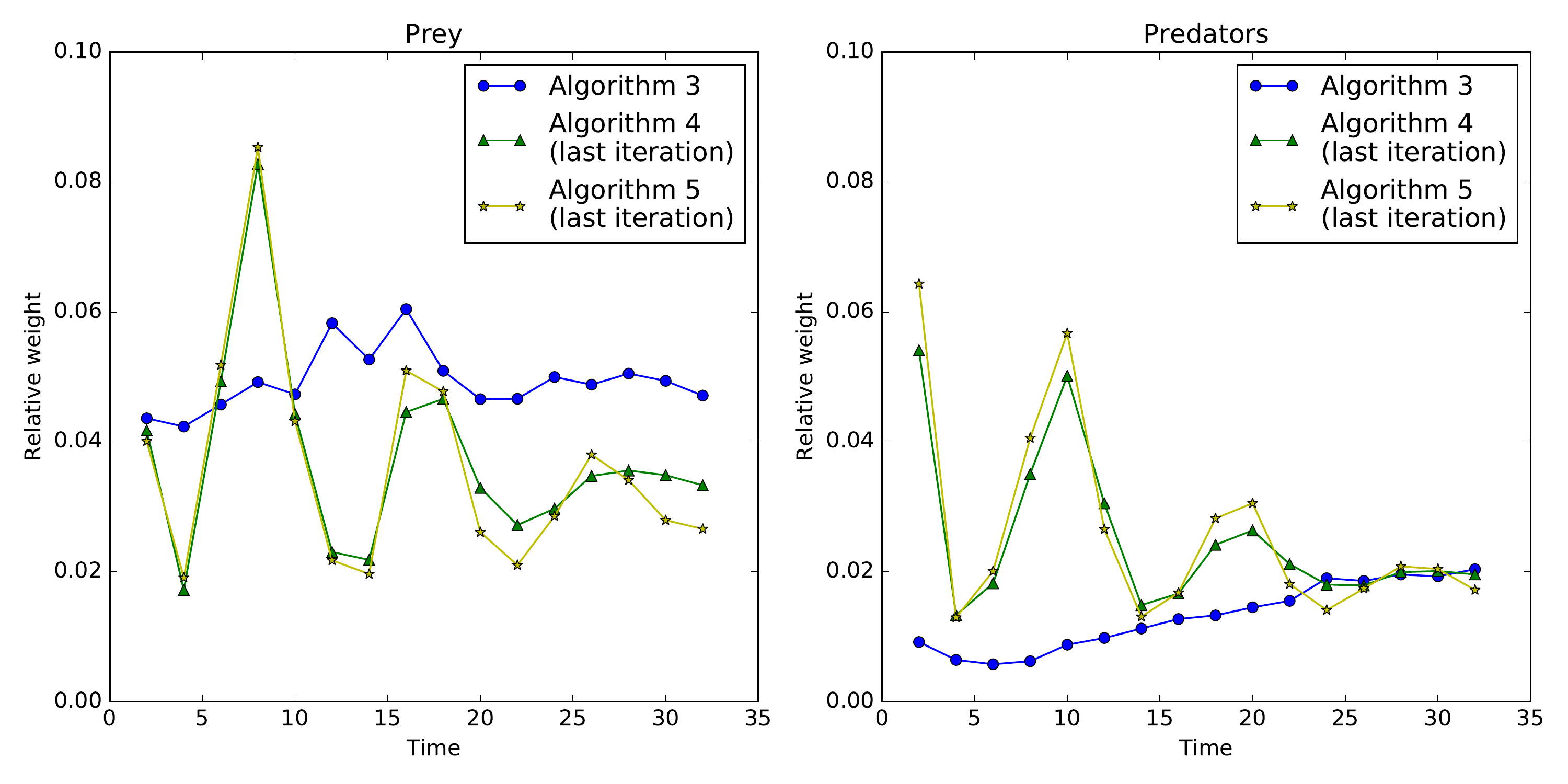}
\caption{Summary statistic weights used in ABC-PMC for the Lotka-Volterra example, rescaled to sum to 1.}
\label{fig:LVweights}
\end{figure*}

\begin{figure*}[pht]
\includegraphics[width=\textwidth]{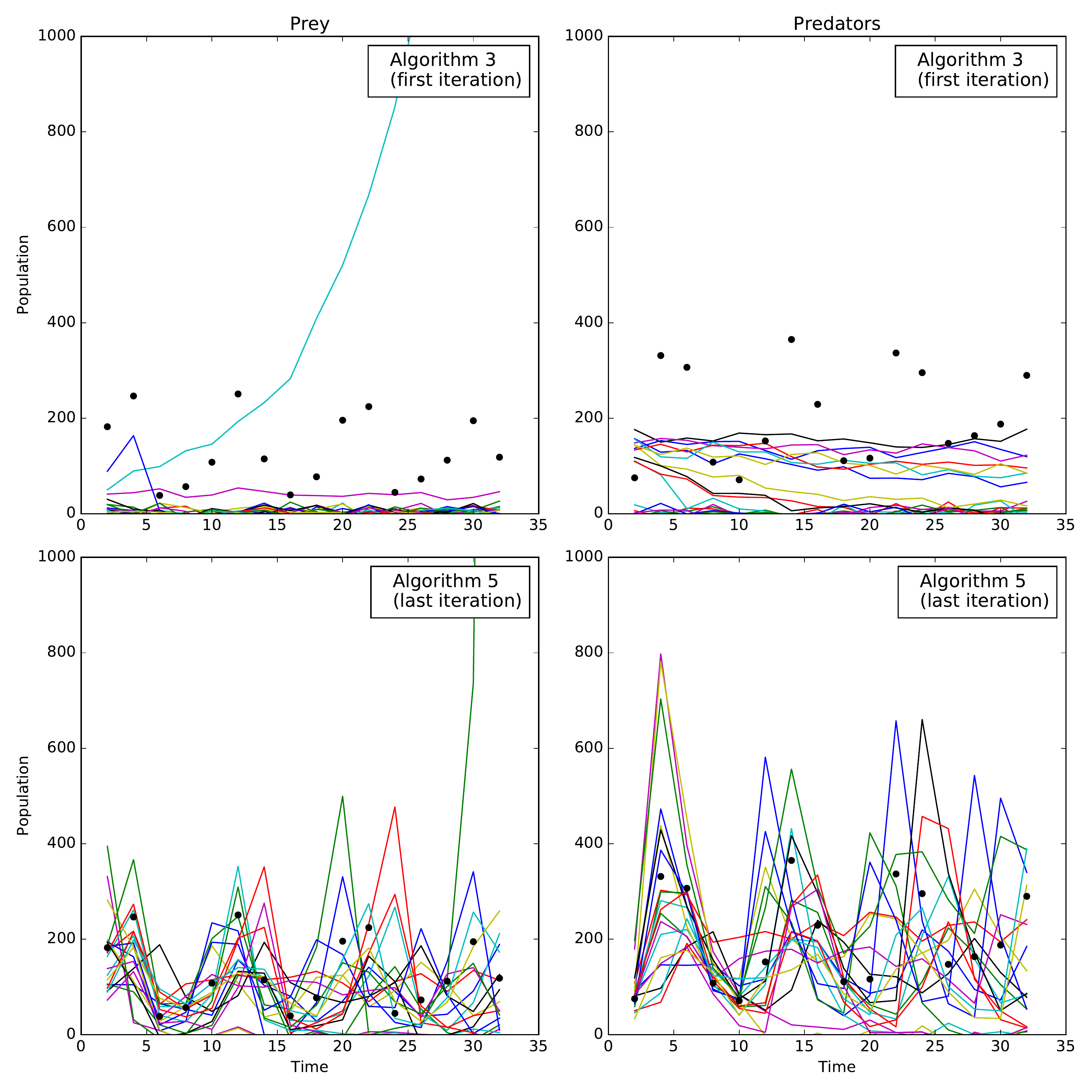}
\caption{Observed dataset (black points) and samples of 20 simulated datasets (coloured lines) for the Lotka-Volterra example.
The top row shows simulations from step 2 of the first iteration of Algorithm \ref{alg:ABCPMC2}.
The bottom row shows simulations from step 2 of the last iteration of Algorithm \ref{alg:adaptive2}.
These are representative examples of the simulations used to select the weights shown in Figure \ref{fig:LVweights}.
Simulations for Algorithm \ref{alg:adaptive1} are not shown but are qualitatively similar to the bottom row.
}
\label{fig:LVpaths}
\end{figure*}

Figure \ref{fig:LVmse} shows the MSEs resulting from the analyses.
The adaptive algorithms, \ref{alg:adaptive1} and \ref{alg:adaptive2}, have similar outputs.
Both produce smaller errors than Algorithm \ref{alg:ABCPMC2} for all parameters after roughly 10,000 simulations.
Table \ref{tab:LVpost} demonstrates that the main difference in the final estimated posteriors is that Algorithm \ref{alg:ABCPMC2} has higher variances.
Figure \ref{fig:LVweights} shows the weights used throughout Algorithm \ref{alg:ABCPMC2} and those used in the final iteration of the others.
Again the adaptive algorithms are similar to each other but different to Algorithm \ref{alg:ABCPMC2}.
Figure \ref{fig:LVpaths} explains this by showing a sample of simulated datasets on which these weights are based.
Under the prior predictive distribution (shown in the top row), at least one population usually quickly becomes extinct,
illustrating that the prior distribution concentrates on the wrong system dynamics and so is unsuitable for choosing distance weights for later iterations of the algorithm.

\begin{table}[hptb]
\begin{center}
\begin{tabular}{c|cccc}
                              & $\log \theta_1$ & $\log \theta_2$ & $\log \theta_3$ \\
\hline
Algorithm \ref{alg:ABCPMC2}   & -0.048 (0.15)   & -5.15 (0.21)    & -0.48 (0.22)    \\
Algorithm \ref{alg:adaptive1} & -0.021 (0.10)   & -5.24 (0.11)    & -0.56 (0.13)    \\
Algorithm \ref{alg:adaptive2} & -0.021 (0.10)   & -5.24 (0.11)    & -0.55 (0.12)
\end{tabular}
\end{center}
\caption{Estimated marginal posterior means and standard deviations (in brackets) of each parameter in the Lotka-Volterra example, for analysis of a particular simulated dataset.
The values are taken from the final iteration of each algorithm.
The true values are $0, -5.30$ and $-0.51$.
(n.b.~All the estimated posteriors are roughly normal.)
} \label{tab:LVpost}
\end{table}

\section{Discussion} \label{sec:conclusion}



This paper has presented two ABC-PMC algorithms with adaptive distance functions.
The algorithms adapt the structure to ABC-PMC by using the output of existing simulation steps to adapt their distance functions.
Therefore they have a similar computational cost for the same number of iterations.
Furthermore, their convergence properties are similar to ABC-PMC.
Several examples have been shown where the new algorithms improve performance.
This is because in each example the scale of the summary statistics varies significantly between prior and posterior predictive distributions.
Of the two algorithms, Algorithm \ref{alg:adaptive1} is simpler to implement, involving only a small modification to standard ABC-PMC, and has essentially the same performance to Algorithm \ref{alg:adaptive2} in two of the three examples.
Algorithm \ref{alg:adaptive2} performs better in the example of Section \ref{sec:illustration_normal}, suggesting it is preferable in situation where continual adaptation is required.
The remainder of this section discusses possibilities to extend this work.

Several variations on ABC-PMC have been proposed in the literature.
The adaptive distance function idea introduced here can be used in most of these.
This is particularly simple for ABC model choice algorithms \citep[e.g.][]{Toni:2009}.
Here, instead of proposing $\theta^*$ values from an importance density,
$(m^*, \theta^*)$ pairs are proposed, where $m^*$ is a model indicator.
This could be implemented in Algorithms \ref{alg:adaptive1} and \ref{alg:adaptive2} while leaving the other details unchanged.
\cite{Drovandi:2011macroparasite}, \cite{delMoral:2012} and \cite{Lenormand:2013} propose ABC-SMC algorithms which update the population of $(\theta, \s)$ pairs between iterations in different ways to ABC-PMC.
In all of these it seems possible to update distance functions using the strategies of Algorithms \ref{alg:adaptive1} and \ref{alg:adaptive2}.
However some of these variations would require additional convergence results to those given in Appendix \ref{sec:convergence}.

Several aspects of Algorithms \ref{alg:adaptive1} and \ref{alg:adaptive2} could be modified.
One natural alternative is to use Mahalanobis-style distance functions $d^t(\x,\y)=\left[ (\x-\y)^T W^t (\x-\y) \right]^{1/2}$ where $W^t$ is an estimate of the precision matrix.
Scenarios exist in which this performs much better than weighted Euclidean distance, \eqref{eq:weighted_d} (Sisson, personal communication).
However exploratory work found it gave similar or worse performance for the examples in this paper.
Distance \eqref{eq:weighted_d} is preferred here for this reason,
and also because its weights are easier to interpret and there are more potential numerical difficulties in estimating a precision matrix.
Nonetheless, for other problems it may be worth considering both alternatives.


Another reason it may be desirable to modify the distance function \eqref{eq:weighted_d} is if some summary statistic, say $s_i$, has an observed value far from most simulated values.
In this case $|s_{\text{obs},i} - s_i|$ can be much larger than $\sigma_i$,
and so $s_i$ can dominate the distances used in this paper.
It is tempting to downweight $s_i$ so that the others summaries can also contribute.
Finding a good way to do this without ignoring $s_i$ altogether is left for future work.

Algorithms \ref{alg:adaptive1} and \ref{alg:adaptive2} update the distance function at each iteration.
There may be scope for similarly updating other tuning choices.
It is particularly appealing to try to improve the choice of summary statistics as the algorithm progresses (as suggested by \citealp{Barnes:2012}.)
Summary statistics could be selected at the same time as the distance function based on the same simulations, for example by a modification of the regression method of \cite{Fearnhead:2012}.
Further work would be required to ensure the convergence of such an algorithm.



\paragraph{Acknowledgements}
Thanks to Michael Stumpf and Scott Sisson for helpful discussions, and three anonymous referees for feedback including suggesting Algorithm \ref{alg:adaptive1}.
The main part of this work was completed while the author was supported by a Richard Rado postdoctoral fellowship from the University of Reading. 
\appendix

\section{Convergence of ABC-PMC algorithms} \label{sec:convergence}


Algorithm \ref{alg:ABCIS} is an ABC importance sampling algorithm.
This appendix considers a sequence of these algorithms.
Denote the acceptance threshold and distance function in the $t$th element of this sequence as $h_t$ and $d^t(\cdot,\cdot)$.
The ABC-PMC algorithms in this paper can be viewed as sequences of this form with specific choices of how $h_t$ and $d^t$ are selected.
Note ABC-rejection is a special case of Algorithm \ref{alg:ABCIS} with $q(\theta)=\pi(\theta)$,
so this framework can also investigate its convergence as $h \to 0$.

\begin{Algorithm}[htp]
\caption{ABC importance sampling}
\label{alg:ABCIS}
\begin{enumerate}
\item Sample $\theta^*_i$ from density $q(\theta)$ independently for $1 \leq i \leq N$.
\item Sample $\y^*_i$ from $\pi(\y|\theta^*_i)$ independently for $1 \leq i \leq N$.
\item Calculate $\s^*_i=S(\y^*_i)$ for $1 \leq i \leq N$.
\item Calculate $d^*_i = d(\s^*_i, \sobs)$.
\item Calculate $w^*_i = \pi(\theta^*_i)/q(\theta^*_i)$ (where $\pi(\theta)$ is the prior density)
\item Return $\{ (\theta^*_i, w^*_i) | d^*_i \leq h \}$.
\end{enumerate}
\end{Algorithm}

The output of importance sampling is a weighted sample $(\theta_i, w_i)_{1 \leq i \leq P}$ for some value of $P$.
A Monte Carlo estimate of $E[h(\theta) | \sobs]$ for an arbitrary function $h(\cdot)$ is then $\frac{\sum_{i=1}^P h(\theta)_i w_i}{\sum_{i=1}^P w_i}$.
For large $P$ this asymptotically equals (as shown in \citealp{Prangle:2011} for example) the expectation under the following density:
\[
\pi_{\ABC, t}(\theta | \sobs) \propto \int \pi(\s | \theta)  \pi(\theta) \mathbbm{1}[d_t(\s,\sobs) \leq h_t] d\s,
\]
known as the ABC posterior.

\begin{theorem} \label{thm:convergence}
Under conditions C1-C5, $\lim_{t \to \infty} \pi_{\ABC,t}(\theta | \sobs) = \pi(\theta | \sobs)$ for almost every choice of $(\theta, \sobs)$ (with respect to the density $\pi(\theta,\s)$).
\end{theorem}

The conditions are:
\begin{enumerate}
\item[C1.]
$\theta \in \mathbb{R}^n$, $\s \in \mathbb{R}^m$ for some $m,n$ and these random variables have density $\pi(\theta, \s)$ with respect to Lebesgue measure.
\item[C2.]
The sets $A_t = \{ \s | d_t(\s,\sobs) \leq h_t \}$ are Lebesgue measurable.
\item[C3.]
$\pi(\sobs) > 0$.
\item[C4.]
$\lim_{t \to \infty} |A_t| = 0$ (where $|\cdot|$ represents Lebesgue measure.)
\item[C5.]
The sets $A_t$ have \emph{bounded eccentricity}.
\end{enumerate}

The definition of bounded eccentricity is that for any $A_t$, there exists a set
$B_t = \{ \s \ | \  ||\s-\sobs||_2 \leq r_t \}$
such that $A_t \subseteq B_t$ and $|A_t| \geq c |B_t|$,
where $||.||$ denotes the Euclidean norm and $c>0$ is a constant.

\paragraph{Proof.}
Observe that:

\begin{align*}
\lim_{t \to \infty} \pi_\ABC(\theta | \sobs)
&= \lim_{t \to \infty} \frac{\int \pi(\theta, \s) \mathbbm{1}(\s \in A_t) d\s}{\int \pi(\theta, \s) \mathbbm{1}(\s \in A_t) d\s d\theta} \\
&= \lim_{t \to \infty} \frac{\int_{\s \in A_t} \pi(\theta, \s) d\s}{\int_{\s \in A_t} \pi(\s) d\s} \\
&= \frac{\lim_{t \to \infty} \tfrac{1}{|A_t|} \int_{\s \in A_t} \pi(\theta, \s) d\s}{\lim_{t \to \infty} \tfrac{1}{|A_t|} \int_{\s \in A_t} \pi(\s) d\s} \\
&= \frac{\pi(\theta, \sobs)}{\pi(\sobs)} \quad \text{almost everywhere} \\
&= \pi(\theta | \sobs).
\end{align*}

The fourth equality follows from the Lebesgue differentiation theorem, which requires conditions C4 and C5.
For more details see \cite{Stein:2009} for example.

\bibliography{ABCdistance}

\begin{thebibliography}{}

\bibitem[Barnes et~al., 2012]{Barnes:2012}
Barnes, C.~P., Filippi, S., and Stumpf, M. P.~H. (2012).
\newblock Contribution to the discussion of {Fearnhead} and {Prangle} (2012).
\newblock {\em {Journal of the Royal Statistical Society: Series B}}, 74:453.

\bibitem[Beaumont, 2010]{Beaumont:2010}
Beaumont, M.~A. (2010).
\newblock Approximate {B}ayesian computation in evolution and ecology.
\newblock {\em Annual Review of Ecology, Evolution and Systematics},
  41:379--406.

\bibitem[Beaumont et~al., 2009]{Beaumont:2009}
Beaumont, M.~A., Cornuet, J.-M., Marin, J.-M., and Robert, C.~P. (2009).
\newblock Adaptive approximate {B}ayesian computation.
\newblock {\em Biometrika}, pages 2025--2035.

\bibitem[Beaumont et~al., 2002]{Beaumont:2002}
Beaumont, M.~A., Zhang, W., and Balding, D.~J. (2002).
\newblock Approximate {B}ayesian computation in population genetics.
\newblock {\em Genetics}, 162:2025--2035.

\bibitem[Bezanson et~al., 2012]{Bezanson:2012}
Bezanson, J., Karpinski, S., Shah, V.~B., and Edelman, A. (2012).
\newblock Julia: A fast dynamic language for technical computing.
\newblock {\em arXiv preprint arXiv:1209.5145}.

\bibitem[Biau et~al., 2015]{Biau:2015}
Biau, G., C{\'e}rou, F., and Guyader, A. (2015).
\newblock New insights into approximate {B}ayesian computation.
\newblock {\em Annales de l'Institut Henri Poincar{\'e} (B) Probabilit{\'e}s et
  Statistiques}, 51(1):376--403.

\bibitem[Blum et~al., 2013]{Blum:2013}
Blum, M. G.~B., Nunes, M.~A., Prangle, D., and Sisson, S.~A. (2013).
\newblock A comparative review of dimension reduction methods in approximate
  {B}ayesian computation.
\newblock {\em Statistical Science}, 28:189--208.

\bibitem[Bonassi and West, 2015]{Bonassi:2015}
Bonassi, F.~V. and West, M. (2015).
\newblock Sequential {Monte Carlo} with adaptive weights for approximate
  {B}ayesian computation.
\newblock {\em Bayesian Analysis}, 10(1):171--187.

\bibitem[Capp{\'e} et~al., 2004]{Cappe:2004}
Capp{\'e}, O., Guillin, A., Marin, J.-M., and Robert, C.~P. (2004).
\newblock Population {Monte Carlo}.
\newblock {\em Journal of Computational and Graphical Statistics}, 13(4).

\bibitem[Csill{\'e}ry et~al., 2010]{Csillery:2010}
Csill{\'e}ry, K., Blum, M. G.~B., Gaggiotti, O., and Fran{\c{c}}ois, O. (2010).
\newblock Approximate {B}ayesian computation in practice.
\newblock {\em Trends in Ecology {\&} Evolution}, 25:410--418.

\bibitem[Csill{\'e}ry et~al., 2012]{Csillery:2012}
Csill{\'e}ry, K., Fran{\c{c}}ois, O., and Blum, M. G.~B. (2012).
\newblock abc: an {R} package for approximate {B}ayesian computation ({ABC}).
\newblock {\em Methods in Ecology and Evolution}, 3:475--479.

\bibitem[Del~Moral et~al., 2012]{delMoral:2012}
Del~Moral, P., Doucet, A., and Jasra, A. (2012).
\newblock An adaptive sequential {Monte Carlo} method for approximate
  {Bayesian} computation.
\newblock {\em Statistics and Computing}, 22(5):1009--1020.

\bibitem[Drovandi and Pettitt, 2011a]{Drovandi:2011macroparasite}
Drovandi, C.~C. and Pettitt, A.~N. (2011a).
\newblock Estimation of parameters for macroparasite population evolution using
  approximate {B}ayesian computation.
\newblock {\em Biometrics}, 67(1):225--233.

\bibitem[Drovandi and Pettitt, 2011b]{Drovandi:2011replenishment}
Drovandi, C.~C. and Pettitt, A.~N. (2011b).
\newblock Likelihood-free {B}ayesian estimation of multivariate quantile
  distributions.
\newblock {\em Computational Statistics \& Data Analysis}, 55(9):2541--2556.

\bibitem[Fasiolo and Wood, 2015]{Fasiolo:2015}
Fasiolo, M. and Wood, S.~N. (2015).
\newblock Approximate methods for dynamic ecological models.
\newblock {\em arXiv preprint arXiv:1511.02644}.

\bibitem[Fearnhead and Prangle, 2012]{Fearnhead:2012}
Fearnhead, P. and Prangle, D. (2012).
\newblock Constructing summary statistics for approximate {B}ayesian
  computation: Semi-automatic {ABC}.
\newblock {\em Journal of the Royal Statistical Society, Series B},
  74:419--474.

\bibitem[Lenormand et~al., 2013]{Lenormand:2013}
Lenormand, M., Jabot, F., and Deffuant, G. (2013).
\newblock Adaptive approximate {Bayesian} computation for complex models.
\newblock {\em Computational Statistics}, 28(6):2777--2796.

\bibitem[Marin et~al., 2012]{Marin:2012}
Marin, J.-M., Pudlo, P., Robert, C.~P., and Ryder, R.~J. (2012).
\newblock Approximate {B}ayesian computational methods.
\newblock {\em Statistics and Computing}, 22(6):1167--1180.

\bibitem[McKinley et~al., 2009]{McKinley:2009}
McKinley, T., Cook, A.~R., and Deardon, R. (2009).
\newblock Inference in epidemic models without likelihoods.
\newblock {\em The International Journal of Biostatistics}, 5(1).

\bibitem[Owen et~al., 2015]{Owen:2015}
Owen, J., Wilkinson, D.~J., and Gillespie, C.~S. (2015).
\newblock Likelihood free inference for {M}arkov processes: a comparison.
\newblock {\em Statistical applications in genetics and molecular biology},
  14(2):189--209.

\bibitem[Prangle, 2011]{Prangle:2011}
Prangle, D. ({2011}).
\newblock {\em Summary statistics and sequential methods for approximate
  {B}ayesian computation}.
\newblock PhD thesis, {Lancaster University}.

\bibitem[Rayner and MacGillivray, 2002]{Rayner:2002}
Rayner, G.~D. and MacGillivray, H.~L. (2002).
\newblock Numerical maximum likelihood estimation for the g-and-k and
  generalized g-and-h distributions.
\newblock {\em Statistics and Computing}, 12(1):57--75.

\bibitem[Sedki et~al., 2012]{Sedki:2012}
Sedki, M., Pudlo, P., Marin, J.-M., Robert, C.~P., and Cornuet, J.-M. (2012).
\newblock Efficient learning in {ABC} algorithms.
\newblock {\em arXiv preprint arXiv:1210.1388}.

\bibitem[Silk et~al., 2013]{Silk:2013}
Silk, D., Filippi, S., and Stumpf, M. P.~H. (2013).
\newblock Optimizing threshold-schedules for sequential approximate {B}ayesian
  computation: applications to molecular systems.
\newblock {\em Statistical applications in genetics and molecular biology},
  12(5):603--618.

\bibitem[Sisson et~al., 2009]{Sisson:2009}
Sisson, S.~A., Fan, Y., and Tanaka, M.~M. (2009).
\newblock Correction: {Sequential Monte Carlo} without likelihoods.
\newblock {\em Proceedings of the National Academy of Sciences},
  106(39):16889--16890.

\bibitem[Stein and Shakarchi, 2009]{Stein:2009}
Stein, E.~M. and Shakarchi, R. (2009).
\newblock {\em Real analysis: measure theory, integration, and {H}ilbert
  spaces}.
\newblock Princeton University Press.

\bibitem[Toni et~al., 2009]{Toni:2009}
Toni, T., Welch, D., Strelkowa, N., Ipsen, A., and Stumpf, M. ({2009}).
\newblock Approximate {B}ayesian computation scheme for parameter inference and
  model selection in dynamical systems.
\newblock {\em {Journal of The Royal Society Interface}}, {6}({31}):{187--202}.

\end{thebibliography}

\end{document}